\documentclass[twocolumn,showpacs,preprintnumbers,amsmath,amssymb]{revtex4}

\usepackage{graphicx}
\usepackage{dcolumn}
\usepackage{bm}

\begin{document}

\preprint{}

\title{Nonlinear optical response in two-dimensional Mott insulators}

\author{Makoto Takahashi, Takami Tohyama and Sadamichi Maekawa}
 \address{Institute for Materials Reserch, Tohoku University, Sendai 980-8577, Japan}

\date{\today}

\begin{abstract}
We study the third-order nonlinear optical susceptibility $\chi^{(3)}$ and photoexcited states of two-dimensional (2D) Mott insulators by using an effective model in the strong-coupling limit of a half-filled Hubbard model.  In the numerically exact diagonalization calculations on finite-size clusters, we find that the coupling of charge and spin degrees of freedom plays a crucial role in the distribution of the dipole-allowed states with odd parity and the dipole-forbidden states with even parity in the photoexcited states.  This is in contrast with the photoexcited states in one dimension, where the charge and spin degrees of freedom are decoupled.  In the third-harmonic generation (THG) spectrum, the main contribution is found to come from the process of three-photon resonance associated with the odd-parity states.  As a result, the two-photon resonance process is less pronounced in the THG spectrum.  The calculated THG spectrum is compared with recent experimental data.  We also find that $\chi^{(3)}$ with a cross-polarized configuration of pump and probe photons shows spectral distributions similar to $\chi^{(3)}$ with a copolarized configuration, although the weight is small.  These findings will help analyses of the experimental data of $\chi^{(3)}$ in 2D Mott insulators.
\end{abstract}


\maketitle
\section{Introduction}
The charge gap in Mott insulators is a consequence of a strong electron correlation represented by a large on-site Coulomb interaction.  The correlation induces novel phenomena in terms of the interplay of charge and spin degrees of freedom.~\cite{Maekawa}  In one-dimensional (1D) Mott insulators, two particles created by photoexcitation---i.e., an unoccupied site and a doubly occupied site of electrons---can move inside the system without being disturbed by surrounding spins in the background.  This is a manifestation of a separation of the charge and spin degrees of freedom, called the spin-charge separation inherent in 1D correlated electron systems.  Optical responses in the 1D Mott insulators are characterized by this phenomenon.~\cite{Stephan,Gebhard,Jeckelmann,Essler,Mizuno,Kishida1}  In two-dimensional (2D) Mott insulators, on the other hand, the two particles are expected not to be free from the spin degree of freedom, because the propagation of a carrier is known to induce a spin cloud around the carrier as a consequence of the misaligned spins along the carrier-hopping paths.~\cite{Maekawa}  Such an interplay of spin and charge is also one of main subjects of the study of high-temperature superconductivity.

The nature of the two particles in the photoexcited states of the Mott insulators is obtained by examining the linear susceptibility $\chi^{(1)}$ with respect to the applied electric field, which provides information on the dipole-allowed states with odd parity among the photoexcited states.  In addition to $\chi^{(1)}$, the third-order nonlinear optical susceptibility $\chi^{(3)}$ is useful to detect not only the odd-parity states but also the dipole-forbidden states with even parity.~\cite{Butcher}  Recently, large values of $\chi^{(3)}$ have been reported for 1D Mott insulators of copper oxides such as Sr$_2$CuO$_3$ ( Ref. 9 and 10 ) from an electro-reflectance measurement~\cite{Kishida2} and pump and probe spectroscopy.~\cite{Ogasawara}  Analyses of $\chi^{(3)}$ have suggested that odd- and even-parity states are nearly degenerate with a large transition dipole moment between them.  Theoretically, $\chi^{(3)}$ in 1D Mott insulators has been examined by employing the numerically exact diagonalization technique for small clusters of the Hubbard model at half filling.~\cite{Mizuno}  It has been shown that odd- and even-parity states are almost degenerate in the same energy region and that the degeneracy is due to the spin-charge separation and strong on-site Coulomb interaction.~\cite{Tohyama}  From this viewpoint, an effective model that can describe the optical nonlinearity has been proposed,~\cite{Mizuno} that is, a holon-doublon model, where holon and doublon represent the charge degree of freedom for photoinduced unoccupied and doubly occupied sites, respectively.  The model reproduces very well the characteristic behaviors of the experimental $\chi^{(3)}$ including data from third-harmonic generation (THG) spectroscopy.~\cite{Kishida1}

In the 2D Mott insulators of copper oxide such as Sr$_2$CuO$_2$Cl$_2$,  $\chi^{(3)}$ has been reported to be one order of magnitude smaller than that in 1D.~\cite{Ashida}  Theoretical calculations based on the (extended) Hubbard model have shown such a dimensionality dependence of $\chi^{(3)}$.~\cite{Mizuno,Ashida}  However, a complete understanding of the nonlinear optical responses in 2D Mott insulators has not been obtained both experimentally and theoretically.  For instance, THG spectra in 2D have been reported from three groups,~\cite{Schulzgen,Schumacher,Kishida3} but the data have not converged: One of the data in La$_2$CuO$_4$ shows several pronounced peak structures in the energy region of 0.6 -- 1.1~eV,\cite{Schulzgen} while other data in La$_2$CuO$_4$ (Ref. 14) and Sr$_2$CuO$_2$Cl$_2$ (Ref. 15) exhibit a broad structure with a maximum at around 0.7~eV.  Although the former data of THG were analyzed by using an excitonic cluster model~\cite{Hanamura} taking into account both copper and oxygen ions, a microscopic description and understanding of photoexcited states in 2D are not complete in terms of the interplay of charge and spin degrees of freedom.

In this paper, we theoretically examine photoexcited states and nonlinear optical responses in the 2D Mott insulators.  We use an effective Hamiltonian of a half-filled Hubbard model in the strong-coupling limit and a numerically exact diagonalization method on finite-size clusters for the calculation of $\chi^{(3)}$.  By using the effective Hamiltnian, we can treat the clusters larger in size than those used in a previous study for the 2D Hubbard model in Ref. 4, leading to new insights into the photoexcited states in the 2D Mott insulators. It is found that the edge of the distribution of the even-parity states is located lower in energy than that of the odd-parity states.  This is different from the distribution in 1D, where the edges of both states almost coincide.  The origin is attributed to the presence of an exchange interaction between spins.  In other words, the spin degree of freedom plays an important role in the photoexcited states.  In the THG spectrum, dominant contributions come from the process of three-photon resonance associated with the odd-parity states.  The two-photon resonance process is hidden by the dominant contributions.  The spectrum obtained by using realistic parameters for 2D copper oxides thus shows broad maxima coming from three-photon resonance.  Such broad spectral features are qualitatively in agreement with the experimental data in Refs. 12 and 13.  We also demonstrate that $\chi^{(3)}$ obtained by a cross-polarized configuration of pump and probe beams shows spectral distributions similar to $\chi^{(3)}$ with a copolarized configuration but with small weight.  These findings will help the analysis of the experimental data of $\chi^{(3)}$ in the 2D Mott insulators.

The rest of this paper is organized as follows.  We introduce an effective Hamiltonian of the half-filled Hubbard model in the strong-coupling limit and show outlines of the procedure to calculate $\chi^{(1)}$ and $\chi^{(3)}$ in Sec.~II.  In Sec.~III, calculated results of the linear and nonlinear optical responses are presented.  The distributions of photoexcited states are discussed in terms of the effect of the spin degree of freedom.  Two-photon absorption (TPA) and THG spectra are calculated with both copolarized and cross-polarized configurations.  We also compare our results with existing experimental data of $\chi^{(3)}$.  The summary is given in Sec.~IV.

\section{\label{sec:level1}model and Method}
The insulating cuprates are known to be charge-transfer- (CT-) type Mott insulators, where both 3$d$ and 2$p$ orbitals exist in the transition-metal and ligand ions, respectively, and participate in the electronic states.  The value of the gap is predominantly determined by the energy position of the 2$p$ orbitals.  However, it is well established that the electronic states of the CT-type insulators can be described by an extended Hubbard model with a half-filled single band by mapping a bound state called the Zhang-Rice singlet state onto the lower Hubbard band.\cite{Maekawa} The extended Hubbard Hamiltonian in 2D is given by
\begin{eqnarray}
&&H_{\rm Hub}=-t\sum_{\langle i,j\rangle_\mathrm{1st}, \sigma} \left( c_{i,\sigma}^\dagger c_{j,\sigma}+ {\rm H.c.} \right) 
+U\sum_i n_{i,\uparrow}n_{i,\downarrow} 
 \nonumber \\
&&-t'\sum_{\langle i,j\rangle_\mathrm{2nd}, \sigma} \left( c_{i,\sigma}^\dagger c_{j,\sigma}+ {\rm H.c.} \right) 
+ V \sum_{\langle i,j\rangle_\mathrm{1st}}n_{i}n_{j},
\label{Hubbard}
\end{eqnarray}
where $c_{i,\sigma}^\dagger$ is the creation operator of an electron with spin $\sigma$ at site $i$, $n_{i,\sigma}=c_{i,\sigma}^{\dagger}c_{i,\sigma}$ and $n_i$=$n_{i,\uparrow}$+$n_{i,\downarrow}$, $\langle i,j\rangle_\mathrm{1st}$ runs over pairs on the nearest-neighbor (NN) sites, $\langle i,j\rangle_\mathrm{2nd}$ runs over pairs on the next-nearest-neighbor (NNN) sites, $t$ ($t'$) is the NN (NNN) hopping integral, $U$ is the on-site Coulomb interaction, and $V$ is the Coulomb interaction between the NN sites.  The value of $t$ is estimated to be $t\sim0.35$ eV from analysis of the electronic structures in the cuprates.~\cite{Maekawa} $t'$ has been systematically found to be of negative sign in contrast to the NN hopping amplitude $t$ with positive sign and to have 40$\%$ of its magnitude ($t'/t=-$0.4).~\cite{Maekawa}  The value of $U$ is estimated to be $U=$10$t$ for the gap values to be consistent with experimental ones.  $V$/$t$ is taken to be 1.~\cite{Neudert}

In the strong-coupling limit ($U \gg t$), the ground state at half filling has one spin per site; i.e., there is no doubly occupied site.  In this case, the low-energy excitation of a Hubbard model may be described by the Heisenberg model, where the dominant exchange interaction is the NN interaction $J$ given by $4t^2/U$.  The Hamiltonian reads
\begin{equation}
H_0=J\sum_{\langle i,j \rangle_{\rm 1st}}\left({\bf S}_i \cdot {\bf S}_j - \frac{1}{4}n_i n_j \right), \label{heisen}
\end{equation}
where ${\bf S}_i$ is the spin operator with $S=1/2$ at site $i$.
However, the photoexcited states that are created by the light have both one doubly occupied site and one vacant site.  In order to obtain an effective Hamiltonian of the extended Hubbard model that describes the photoexcited states, we introduce projection operators $\Pi_0$, $\Pi_1$, and $\Pi_2$ onto the Hilbert space with no doubly occupied site, one doubly occupied site, and two doubly occupied sites, respectively.  The Hubbard Hamiltonian, Eq. (\ref{Hubbard}), is conveniently split into two parts: kinetic terms ($H_t$ and $H_{t'}$) and interacting terms ($H_U$ and $H_V$).  We perform a second-order perturbation with respect to $H_t$.  Since the value of $|t'|$ is smaller than $t$ ($t'=-0.4t$), we neglect $H_{t'}$ in the perturbation process.  The resulting effective Hamiltonian is given by
\begin{equation}
H_{\rm eff}=H_1+H_2+H_3+H_4+U,
\label{Heff}
\end{equation}
where
\begin{widetext}
\begin{eqnarray}
H_1&=&\Pi_1(H_t+H_{t'})\Pi_1 \nonumber \\
 &=&-t \sum_{\langle i,j \rangle_{\rm 1st},\sigma}\left\{(1-n_{i-\sigma})c^{\dagger}_{i\sigma}c_{j\sigma}(1-n_{j-\sigma})+n_{i-\sigma}c^{\dagger}_{i\sigma}c_{j\sigma}n_{j-\sigma} + {\rm H.c.} \right\} \nonumber \\
 & &-t' \sum_{\langle i,j \rangle_{\rm 2nd},\sigma}\left\{(1-n_{i-\sigma})c^{\dagger}_{i\sigma}c_{j\sigma}(1-n_{j-\sigma})+n_{i-\sigma}c^{\dagger}_{i\sigma}c_{j\sigma}n_{j-\sigma} + {\rm H.c.} \right\},\\
H_2&=&-\frac{1}{U}\Pi_1H_t\Pi_2H_t\Pi_1 \nonumber \\
&=&J\sum_{\langle  i,j \rangle_{\rm 1st},\sigma} \left( {\bf S}_{i} \cdot {\bf S}_{j} -\frac{1}{4}n_{i} n_{j} \right) + \frac{t^2}{U}\sum_{i,j,\sigma} \left\{n_{j\sigma}n_{j-\sigma}(1-n_{i-\sigma})(1-n_{i\sigma})+c^{\dagger}_{i \sigma}c^{\dagger}_{i -\sigma}c_{j -\sigma}c_{j\sigma} \right\}  \nonumber \\
&\;&+\frac{t^2}{U}\sum_{i,j,k,\sigma,\sigma '} \{(1-n_{i-\sigma})c^{\dagger}_{i\sigma}c_{j\sigma}n_{j-\sigma}n_{j-\sigma '}c^{\dagger}_{j \sigma '}c_{k \sigma '}(1-n_{k -\sigma '}) \nonumber \\
&\;&\;\;\;\;+(1-n_{j-\sigma})c^{\dagger}_{j\sigma}c_{i \sigma}n_{i -\sigma}n_{k -\sigma '}c^{\dagger}_{k \sigma '}c_{j \sigma '}(1-n_{j -\sigma '}) \} \label{3-sites}, \\
H_3&=&\frac{1}{U}\Pi_1H_t\Pi_0H_t\Pi_1 \nonumber \\
&=&\frac{t^2}{U}\sum_{i,j,\sigma} \{n_{i -\sigma}c^{\dagger}_{i \sigma}c_{k \sigma}n_{k-\sigma}(1-n_{j-\sigma})(1-n_{j\sigma})+c^{\dagger}_{j -\sigma}c_{i -\sigma}c^{\dagger}_{j\sigma}(1-n_{j \sigma})c_{k \sigma}n_{k-\sigma} \nonumber \\
&\;& +n_{i\sigma}c^{\dagger}_{i -\sigma}c^{\dagger}_{k \sigma}(1-n_{k -\sigma})c_{j\sigma}c_{j -\sigma} -(1-n_{k -\sigma})c^{\dagger}_{k \sigma}c_{k \sigma}n_{k-\sigma}(1-n_{i-\sigma})n_{j-\sigma}n_{j \sigma} \},
\end{eqnarray}
\end{widetext}
and
\begin{eqnarray}
H_4&=&-V\sum_{i,j}n_{i\uparrow}n_{i\downarrow}(1-n_{j\uparrow})(1-n_{j\downarrow}).
\label{H4}
\end{eqnarray}
Here, $H_1$ expresses hopping process of unoccupied and doubly occupied sites.  $H_2$ contains the exchange interaction and so-called three-site terms.  $H_1$ and $H_2$ are similar to the $t$-$J$ model with three-site terms and NNN hopping, while $H_3$ has specific terms of the photoexcited states.  The attractive interaction between the unoccupied and doubly occupied sites is described by $H_4$.  

The electric field $\bf E$ of the incident light induces the dielectric polarization $\bf P$ in a material, which is described by a power series of nonlinear optical susceptibility $\chi ^{(n)}$:  $\mathbf{P}=\epsilon_0(\chi^{(1)}\mathbf{E}+\chi^{(2)}\mathbf{E}^2+\chi^{(3)}\mathbf{E}^3+\cdots)$.
The linear susceptibility $\chi^{(1)}$ is given by
\begin{eqnarray}
&& \chi^{(1)}_{kl} (-\omega; \omega) \nonumber \\
&&=\frac{1}{\epsilon_0 L} \frac{e^2}{\hbar}
\sum_a \left ( \frac{r_{0a}^k r_{a0}^l}{\Omega_a-i\Gamma_a-\omega} + \frac{r_{0a}^l r_{a0}^k}{\Omega_a+i\Gamma_a+\omega} \right ) ,
\label{chi1}
\end{eqnarray}
where $\epsilon_0$ is the dielectric constant, $k$ and $l$ are the polarization directions and $er_{0a} (=e\langle 0 | \hat{\bf r} | a \rangle) $ is the dipole moment between the ground state $|0 \rangle$ of the Heisenberg model (\ref{heisen}) and photoexcited state $|a \rangle$ with odd parity obtained from the effective Hamiltonian (\ref{Heff}), $\hat{\bf r}$ being the dipole displacement operator.  $\Omega_a$($=E_a-E_0$) is the energy difference between $|0 \rangle $ and $|a \rangle$, $L$ is the number of sites, and $\Gamma_a$ is the damping factor. The distribution of odd-parity states among the photoexcited states is obtained by examining the imaginary part of $\chi^{(1)}_{ll}$ in the $l$ polarization direction, which is related to the dynamical current-current correlation function $\alpha_{ll} (\omega) $ as Im $\chi^{(1)}_{ll}(-\omega;\omega)=\alpha_{ll}(\omega)/\omega^2$.  Here $\alpha_{ll}(\omega)$ reads
\begin{equation}
\alpha_{ll}(\omega) = \frac{\pi}{\epsilon_0 L }\frac{e^2}{\hbar} \sum_{a} |\langle a| \hat{\bf j}_l| 0 \rangle |^2 \delta(\omega - E_a + E_0) ,
\end{equation}
where $\hat{\bf j}$$_l$ is the $l$ component of the current operator $\hat{\bf j}$ which is given by 
\begin{eqnarray}
\hat{\bf j}&=&i\{t\sum_{\langle i,j\rangle_\mathrm{1st}, \sigma} ({\bf R}_j-{\bf R}_i)(c_{i,\sigma}^{\dagger}c_{j,\sigma}-c_{j,\sigma}^{\dagger}c_{i,\sigma}) \nonumber \\
&&+t'\sum_{\langle i,j\rangle_\mathrm{2nd}, \sigma}({\bf R}_j-{\bf R}_i)(c_{i,\sigma}^{\dagger}c_{j,\sigma}-c_{j,\sigma}^{\dagger}c_{i,\sigma})\}.
\end{eqnarray}
Here, ${\bf R}_i$ is the position vector at site $i$.  The distribution of even-parity states can be detected by a correlation function $\beta_{ll} (\omega)$ in which the current operator $\hat{\bf j}_l$ in $\alpha_{ll}(\omega)$ is replaced by the $l$ component of the stress tensor operator $\hat{\mathbf{\tau}}$,
\begin{equation}
\beta_{ll}(\omega) = \frac{\pi}{\epsilon_0 L }\frac{e^2}{\hbar} \sum_{b} |\langle b| {\bf \hat{\tau}}_l| 0 \rangle |^2 \delta(\omega - E_b + E_0) ,
\end{equation}
with
\begin{eqnarray}
\hat{\bf \tau}&=&t\sum_{\langle i,j\rangle_\mathrm{1st}, \sigma} ({\bf R}_j-{\bf R}_i)(c_{i,\sigma}^{\dagger}c_{j,\sigma}+c_{j,\sigma}^{\dagger}c_{i,\sigma}) \nonumber \\
&+&t'\sum_{\langle i,j\rangle_\mathrm{2nd}, \sigma}({\bf R}_j-{\bf R}_i)(c_{i,\sigma}^{\dagger}c_{j,\sigma}+c_{j,\sigma}^{\dagger}c_{i,\sigma}),
\end{eqnarray}
where $b$ denotes even-parity state $| b \rangle$ with energy $E_b$.  Due to symmetry restrictions, the second-order susceptibility $\chi^{(2)}$ vanishes in centrosymmetric materials to which the insulating cuprates belong.  The third-order susceptibility $\chi^{(3)}$ is expressed as
\begin{widetext}
\begin{eqnarray}
&&\chi^{(3)}_{klmn} (-\omega_\sigma; \omega_1, \omega_2, \omega_3) \nonumber \\
&=&\frac{1}{\epsilon_0 L} \frac{e^4}{3!\hbar^3} {\bf {\cal P}}\sum_{a,b,c}\frac{r_{0a}^k r_{ab}^l r_{bc}^m r_{c0}^n}{(\Omega_a-i\Gamma_a-\omega_\sigma)(\Omega_b-i\Gamma_b-\omega_2-\omega_3)(\Omega_c-i\Gamma_c-\omega_3)} \label{chi3} \\
&&=\frac{1}{\epsilon_0 L} \frac{e^4}{3!\hbar^3} {\bf {\cal P}}\gamma_{klmn} (\omega_{\sigma}+i \Gamma_{a} ,\omega_2+\omega_3+i \Gamma_b ,\omega_3+i \Gamma_c ), \label{eq:gamma}
\end{eqnarray}
\end{widetext}
where $c$ denotes odd-parity states $|c\rangle$ with energy $E_c$, $\omega_\sigma=\omega_1+\omega_2+\omega_3$,  and ${\bf {\cal P}}$ represents the sum of permutation on ($l$,$\omega_1$), ($m$,$\omega_2$), ($n$,$\omega_3$), and ($k$,$\omega_\sigma$).  Hereafter, the damping factors $\Gamma_a$, $\Gamma_b$ and $\Gamma_c$ are assumed to have the same value $\Gamma$, with $\Gamma/t=0.4$, and we take  $e=\hbar=\epsilon_0=1$.  

$\gamma_{klmn}$ in Eq. (\ref{eq:gamma}) is rewritten as
\begin{widetext}
\begin{eqnarray}
&&\gamma_{klmn} (z_p ,z_q ,z_r )=\sum\limits_{a,b,c} {\frac{r^{k}_{0a} r^{l}_{ab} r^{m}_{bc} r^{n}_{c0}}{{(E_a  - E_0  - z_{p})(E_b  - E_0  - z_q)(E_c  - E_0  - z_r)}}} \label{gamma},
\end{eqnarray}
where $z_p=\omega _p+ i \Gamma$.
This quantity can be calculated by using the correction vector technique~\cite{Soos}:
\begin{eqnarray}
\gamma_{klmn} (z_p ,z_q ,z_r )&=& \sum\limits_{a,b,c} {\frac{{\left\langle 0 \right|\hat{{\bf r}}^k\left| a \right\rangle \left\langle a \right|\hat{{\bf r}}^l\left| b \right\rangle \left\langle b \right|\hat{{\bf r}}^m\left| c \right\rangle \left\langle c \right|\hat{{\bf r}}^n\left| 0 \right\rangle }}{{(E_a  - E_0  - z_{p})(E_b  - E_0  - z_q)(E_c  - E_0  - z_r)}}} \label{cv1} \nonumber \\
  &=& \left\langle 0 \left|\hat{{\bf r}}^k\frac{1}{{H_{\rm eff} - E_0  - z_p}}\hat{{\bf r}}^l\frac{1}{{H_{\rm eff} - E_0  - z_q}}\hat{{\bf r}}^m\frac{1}{{H_{\rm eff} - E_0  - z_r}}\hat{{\bf r}}^n\right| 0 \right\rangle \label{cv2} \\ 
  &=& \left\langle {\phi (z_p)} \left|\hat{{\bf r}}^l\frac{1}{{H_{\rm eff} - E_0  - z_q}}\hat{{\bf r}}^m\right| {\phi (z_r)} \right\rangle \label{cv3} \nonumber\\ 
  &=& \left\langle {\phi (z_p)} \right|\hat{{\bf r}^l}\left| {\psi (z_q,z_r)} \right\rangle \label{cv4} \nonumber
\end{eqnarray}
\end{widetext}
where $\hat{{\bf r}}^k$ is the $k$ component of $\hat{\bf r}$.  The correction vectors $\phi(z_r)$ and $\psi(z_q,z_r)$ are obtained by solving the following equations: 
\begin{eqnarray}
(H_{\rm eff} - E_0  - z_r)\left| {\phi (z_r)} \right\rangle  &=& \hat{{\bf r}}^n \left| 0 \right\rangle  \label{relation1}\\
 (H_{\rm eff} - E_0  - z_q)\left| {\psi (z_q,z_r)} \right\rangle  &=& \hat{{\bf r}}^m \left| {\phi (z_r)} \right\rangle , \label{relation2}
\end{eqnarray}
where $\hat{{\bf r}}^n|0\rangle$ is obtained by solving $\hat{{\bf j}}_n|0\rangle=i (H_{\rm eff}-E_0)\hat{{\bf r}}^n|0\rangle$.

In order to describe 2D systems, we employ an 18-site cluster with open boundary condition, which is shown in the inset of Fig. 1(a).  The ground state $|0\rangle$ of this cluster is obtained by applying the standard Lanczos method to the Heisenberg Hamiltonian (\ref{heisen}).  The correlation functions $\alpha_{ll}(\omega)$ and $\beta_{ll}(\omega)$ are also calculated by the Lanczos technique.  $\chi^{(3)}$ is obtained from Eqs. (\ref{cv2})--(\ref{relation2}).  
In the following section, we consider two kinds of $\chi^{(3)}$ spectra: One is theTPA spectrum which is defined as Im$\chi^{(3)}(-\omega;-\omega,\omega,\omega)$, and the other is THG spectrum which is defined as $|\chi^{(3)}(-3\omega;\omega,\omega,\omega)|$.  
Since the 18-site cluster has no 90$^\circ$ rotational symmetry, two spectra in the $x$- and $y$-polarization directions are inequivalent.  However, qualitatevely similar features are obtained between the two spectra.   We have also calculated the $\chi^{(3)}$ spectra by using a 4$\times$4 cluster that preserves the 90$^\circ$ rotational symmetry.  The results are found to be consistent with those of the 18-site cluster, indicating small size effects.  Therefore, in the next section, we take an average of the spectra of the 18-site cluster and use the following notation: Im$\chi^{(1)}(-\omega;\omega)=$\{Im$\chi^{(1)}_{xx}(-\omega;\omega)+$Im$\chi^{(1)}_{yy}(-\omega;\omega)\}/2$, $\alpha(\omega)=\{\alpha_{xx}(\omega)+\alpha_{yy}(\omega)\}/2$, $\beta(\omega)=\{\beta_{xx}(\omega)+\beta_{yy}(\omega)\}/2$, $\chi^{\mathrm{TPA}}_{\parallel}=\{$Im$\chi^{(3)}_{xxxx}(-\omega;-\omega,\omega,\omega)+$ Im$\chi^{(3)}_{yyyy}(-\omega;-\omega,\omega,\omega)\}/2$, and $\chi^{\mathrm{THG}}_{\parallel}=\{|\chi^{(3)}_{xxxx}(-3\omega;\omega,\omega,\omega)|+|\chi^{(3)}_{yyyy}(-3\omega;\omega,\omega,\omega)|\}/2$.  For cross-polarized configurations of the pump and probe photons, we take the following averages: $\chi^{\mathrm{TPA}}_{\perp}=\{$Im$\chi^{(3)}_{xxyy}(-\omega;-\omega,\omega,\omega)+$ Im$\chi^{(3)}_{yyxx}(-\omega;-\omega,\omega,\omega)\}/2$ and $\chi^{\mathrm{THG}}_{\perp}=\{|\chi^{(3)}_{xxyy}(-3\omega;\omega,\omega,\omega)|+|\chi^{(3)}_{yyxx}(-3\omega;\omega,\omega,\omega)|\}/2$.

\section{Results and discussions}
\subsection{Linear absorption and two-photon absorption}
We first show the distribution of optical-allowed states with odd parity and optical-forbidden states with even parity among the photoexcited states created by photoexcitation.  Figure~1(a) exhibits $\alpha(\omega)$, which detects the odd-parity states.  Here $\delta$ functions denoted by vertical bars are broadened by a Lorentzian with a width of 0.4$t$.  There are two broad peaks at $\omega-U\sim-3t$ and $-t$, which is in contrast with 1D where only a single peak appears [see Fig.~3(b)].  We note that the higher-energy peak plays important roles in the resonance condition of two-magnon Raman scattering as will be reported elsewhere.~\cite{Onodera}  The distribution of even-parity states detected by $\beta(\omega)$ is shown in Fig.~1(b).  The lowest-energy broad peak is located at $\omega-U\sim-3.5t$, which is lower than that of $\alpha(\omega)$.  This is also different from the case of 1D where a one-to-one correspondence between the odd- and even-parity eigenstates is clearly seen.~\cite{Mizuno,Kishida1,Tohyama} Such a degeneracy in 1D is a consequence of the spin-charge separation and strong on-site Coulomb interaction.  In 2D, however, motion of charge carriers induces a spin cloud around the carriers as a consequence of the misaligned spins along the carrier-hopping paths.  Therefore, the effect of the coupling of charge and spin should appear in the distribution of odd- and even-parity states.  This effect can be checked  by changing the value of the exchange interaction $J$/$t$.  In Fig. 2, we show $\alpha(\omega)$ and $\beta(\omega)$ with the same parameters as used in Fig. 1 but $U/t=100$ ($J/t=0.04$).  By changing $J/t$ from 0.4 to 0.04, we find that the higher-energy peak at $\omega-U=-t$ in Fig. 1(a) is smoothly connected to a broad peak at $\omega-U=-2.5t$ in Fig. 2(a).  On the other hand, the lowest-energy peak in Fig. 1(a) loses its weight with reducing $J/t$, and only a small amount of the weight is seen around $\omega-U=-5t$ in Fig. 2(a).  Similarly, the lowest-energy peak of $\beta(\omega)$ seen in Fig. 1(b) almost disappears in Fig. 2(b).  Therefore, for small $J/t$, it is hard to identify the difference in  the distribution of the odd- and even-parity states near the edge of the photoexcited states.  This implies that the distribution near the edge is sensitive to the coupling of charge and spin.  In other words, the coupling could be the origin of the lowest-energy peaks in Fig. 1 and also of the lower-energy shift of the even-parity states  as compared with the odd-parity states.  
\begin{figure}
\begin{center}
\includegraphics[width=8.cm]{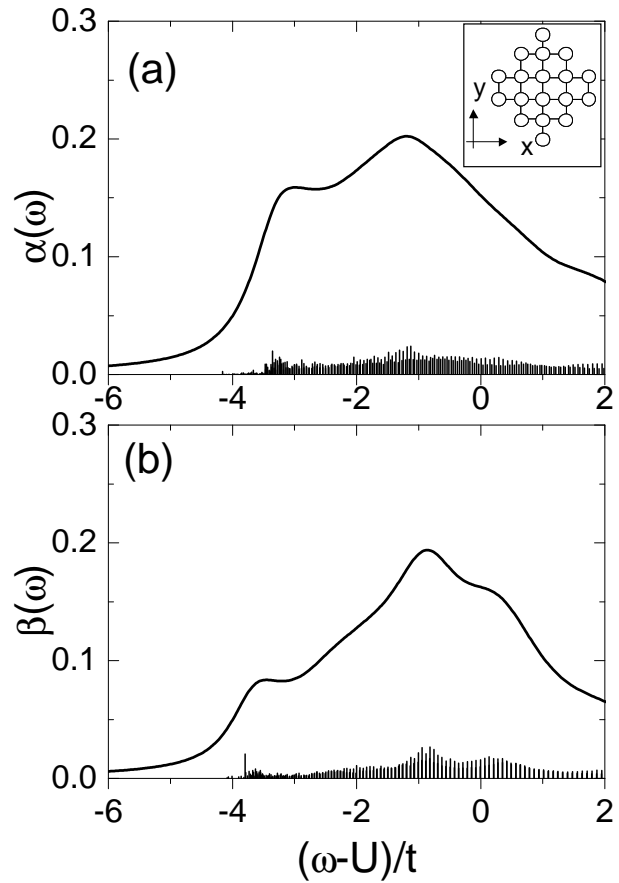}
\caption{\label{fig1} (a) $\alpha (\omega) $ and (b) $\beta (\omega)$ in an 18-site cluster of an effective model in the strong-coupling limit of the Hubbard model with $U/t=10$ $(J/t=0.4)$, $t'/t=-0.4$, and $V/t=1$. The solid lines are obtained by performing a Lorentzian broadening with a width of 0.4$t$ on the $\delta$ functions denoted by vertical bars. The spectra in (a) and (b)  represent the distribution of photoexcited states with odd and even parities, respectively.  The inset in (a) shows the 18-site cluster with open boundary condition.}
\end{center}
\end{figure}
\begin{figure}
\begin{center}
\includegraphics[width=8.cm]{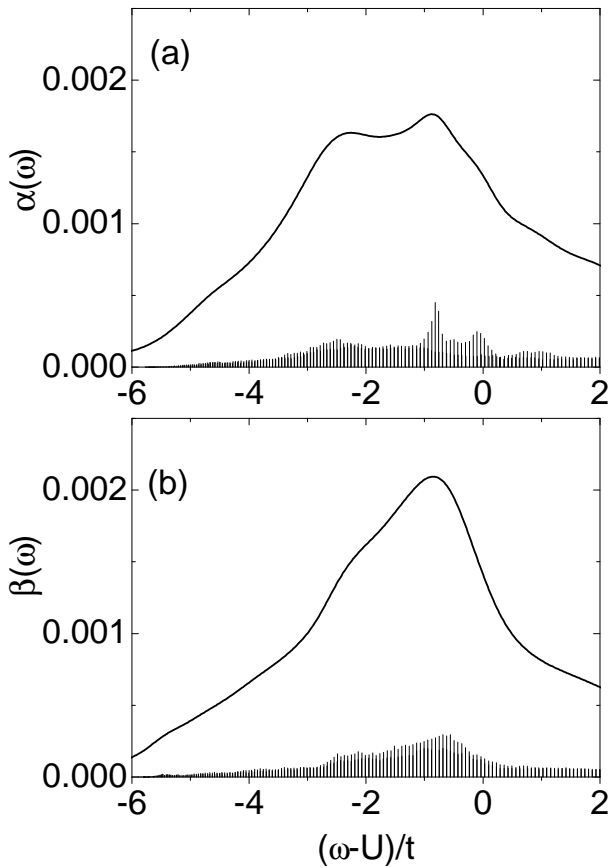}
\caption{\label{fig2} The same as Fig. 1, but $U/t=100$ ($J/t=0.04$). }
\end{center}
\end{figure}
\begin{figure}
\begin{center}
\includegraphics[width=8.5cm]{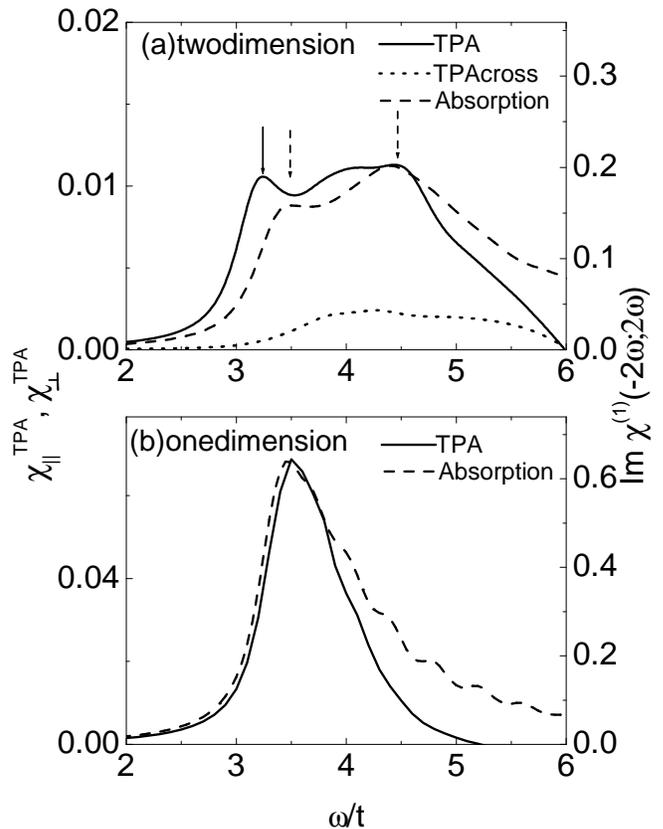}
\caption{\label{fig3} (a) $\chi^{\rm TPA}_{\parallel}$ (solid line), $\chi_{\bot}^{\rm TPA}$ (dotted line), and Im$\chi^{(1)}(-2\omega;2\omega)$ (dashed line) in an 18-site 2D cluster of an effective model in the strong-coupling limit of the Hubbard model with $U/t=10$, $t'/t=-0.4$, and $V/t=1$. The solid and dashed arrows denote the peak position of $\chi_{\parallel}^{\rm TPA}$ and Im$\chi^{(1)}(-2\omega;2\omega)$, respectively.
(b) $\chi_{\parallel}^{\rm TPA}$ (solid line) and Im$\chi^{(1)}(-2\omega;2\omega)$ (dashed line) in an 18-site 1D chain with $U/t=10$, $t'/t=0$, and $V/t=1.5$.
}\end{center}
\end{figure}

The pump and probe spectroscopy is employed to detect even-parity states.~\cite{Ogasawara,Ashida}  As a corresponding quantity, we calculate the TPA spectrum $\chi_{\parallel}^{\mathrm{TPA}}$.  In $\chi^{\rm TPA}_{\parallel}$, a dominant contribution comes from the following term:
\begin{equation}
\frac{1}{6L} \sum_{a,b,c} \frac{x_{0a} x_{ab} x_{bc} x_{c0}+y_{0a} y_{ab} y_{bc} y_{c0}}{(\Omega_a-i\Gamma-\omega)(\Omega_b-i\Gamma-2\omega)(\Omega_c-i\Gamma-\omega)},
\label{TPA}
\end{equation}
where the dipole moment between $|0 \rangle$ and $|b \rangle$ in the $x$($y$) direction is represented by $x_{0a}$($y_{0a}$) instead of $r_{0a}^x$($r_{0a}^y$) in Eq. (\ref{chi3}).  The resonance occurs when $2\omega$ is equal to the energy of eigenstates with even parity.  Figure~3(a) shows $\chi^{\rm TPA}_{\parallel}$ (solid line) together with linear absorption Im $\chi^{(1)}(-2\omega;2\omega)$ (dashed line).  In the figure, the two spectra are plotted by making the largest weight the same height.  In order to examine the effect of dimensionality, results for an 18-site 1D chain are also shown in Fig. 3(b), where parameter values are those of the 1D Mott insulator Sr$_2$CuO$_3$ ($U$/$t$=10, $V$/$t$$=$1.5, and $t'/t=0$).  The maximum value of TPA in 2D is nearly 5 times smaller than that in 1D.  We also find that, while in 1D a peak position as well as an edge position is nearly identical between linear absorption and TPA spectra, in 2D both peak and edge positions are lower in TPA than in linear absorption as expected from Fig.~1.  Such dimensionality dependence is qualitatively consistent with experimental data where a low-energy tail of TPA in Sr$_2$CuO$_2$Cl$_2$ down to 1.5~eV was reported.~\cite{Ashida}

Summarizing the dimensionality dependence of optical responses in the Mott insulators, we find that (i) the linear absorption spectrum in 2D shows two broad peaks in contrast to a single peak in 1D, (ii) the magnitude of the TPA spectrum in 2D is nearly 5 times smaller than that in 1D, and (iii) in 2D the peak and edge positions of the TPA spectrum are lower than those of linear absorption in contrast to the case of 1D, where TPA and linear absorption show almost an identical behavior near the edge.  All of the features are consistent with the experimental data of linear absorption and TPA spectra measured in 1D Sr$_2$CuO$_3$ and 2D Sr$_2$CuO$_2$Cl$_2$.~\cite{Ashida}  This agreement implies that our single-band Hubbard model is suitable for a description of the photoexcited states in both 1D and 2D Mott insulators of copper oxides.  In the next subsection, we will examine THG spectra in 2D, for which three experimental results are now available but exhibit some discrepancies between them.

\subsection{Third-harmonic generation}
The THG spectrum $\chi_{\parallel}^{\rm THG}$ contains information about both odd- and even-parity states via multiphoton resonance processes, since a dominant contribution is expressed as
\begin{equation}
\frac{1}{2L} \sum_{a,b,c} \frac{x_{0a} x_{ab} x_{bc} x_{c0}+y_{0a} y_{ab} y_{bc} y_{c0}}{(\Omega_a-i\Gamma-3\omega)(\Omega_b-i\Gamma-2\omega)(\Omega_c-i\Gamma-\omega)},
\label{THG}
\end{equation}
where the resonance occurs when $3\omega$ (three-photon resonance) and $2\omega$ (two-photon resonance) are equal to the eigenenergies of the odd- and even-parity states, respectively.  Figure~4 shows $\chi^{\mathrm{THG}}_{\parallel}$ (solid line).  The spectrum consists of a broad peak and a hump below $\omega\sim3.5t$.  Since the peak position of $\omega\sim2.3t$ (dashed arrow) coincides with one-third of a peak energy at $\omega\sim6.9t$ in the linear absorption spectrum indicated by a dashed arrow in Fig.~3(a), the origin of the peak can be attributed to three-photon resonance.  On the other hand, there are two possibilities for the origin of the broad hump structure around $\omega\sim2.8t$.  One is again three-photon resonance, and the other is two-photon resonance resonating with even-parity states.  The dashed arrow at $\omega=2.9t$ in Fig.~4 indicates one-third of the second-peak energy in the linear absorption spectrum (8.8$t$) of Fig.~3(a) indicated by a dashed arrow, while the solid arrow at $\omega=3.2t$ indicates the energy of the lowest-energy peak in the TPA spectrum of Fig.~3(a) indicated by a solid arrow.  Since the hump position is closer to the dashed arrow, we regard the hump as a result of three-photon resonance.  A contribution of the even-parity states through two-photon resonance is supposed to overlap with three-photon resonance and thus may be hidden as a background in the THG spectrum.

\begin{figure}
\begin{center}
\includegraphics[width=8.cm]{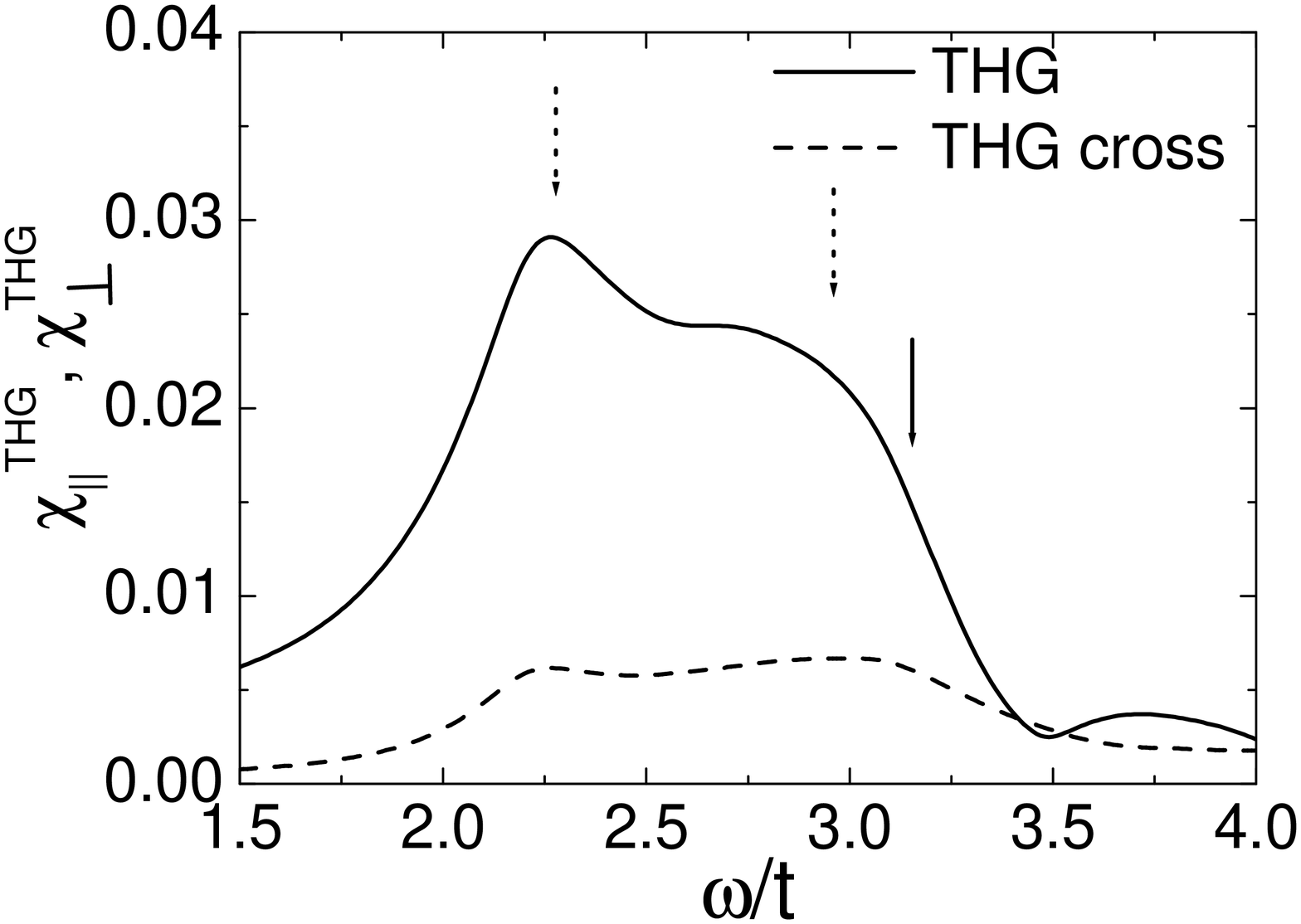}
\caption{\label{fig4} $\chi^{\rm THG}_{\parallel}$ (solid line) and $\chi_{\bot}^{\rm THG}$ (dashed line) in an 18-site 2D cluster of an effective model in the strong-coupling limit of the Hubbard model with $U/t=10$, $t'/t=-0.4$, and $V/t=1$.  The dashed arrows denote the position of three-photon resonance expected from Im$\chi^{(1)}(-2\omega;2\omega)$ in Fig. 3(a).  The solid arrow denotes the position of two-photon resonance expected form $\chi^{\rm TPA}_{\parallel}$.}
\end{center}
\end{figure}
\begin{figure}
\begin{center}
\includegraphics[width=8.cm]{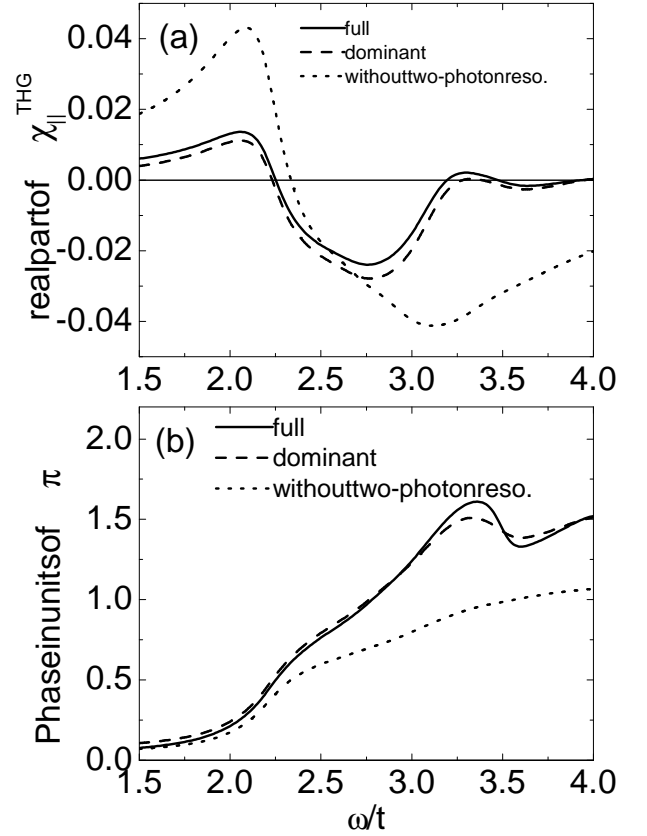}
\caption{\label{fig5} The real part (a) and the phase (b) of $\chi^{\mathrm{THG}}_{\parallel}$ in an 18-site 2D cluster of an effective model in the strong-coupling limit of the Hubbard model with $U/t=10$, $t'/t=-0.4$, and $V/t=1$. The solid lines represent full contribution obtained from Eq. (\ref{chi3}).  The dashed lines represent the contribution from the dominant term, Eq. (\ref{THG}). The dotted lines are obtained by neglecting the two-photon resonance terms in the denominator of Eq. (\ref{THG}).}
\end{center}
\end{figure}

We can check the contribution from the even-parity states to THG, by examining the real part of $\chi^{\mathrm{THG}}_{\parallel}$, which is shown in Fig.~5(a).  The solid curve obtained from Eq.~(\ref{chi3}) changes its sign from negative to positive at $\omega\sim3.2t$, which is close to the two-photon resonance energy (the solid arrow in Fig.~4).  The contribution from the dominant term, Eq.~(\ref{THG}) is also plotted as the dashed curve, which is similar to the solid curve and shows a zero crossing at $\omega\sim3.3t$ . In order to clarify whether such a crossing or sign change is due to a contribution from the even-parity states, we calculate the real part of THG by neglecting the two-photon resonance contribution in the denominator of Eq.~(\ref{THG}) and plot it as the dotted curve in Fig.~5(a).  We find no sign change near the two-photon resonance energy.  Therefore, the sign change is a consequence of the presence of the even-parity states in the THG spectrum.  However, as mentioned above, the weight cannot be recognized as a pronounced structure because of large contributions from the odd-parity states in the same energy region. 

The contribution from the even-parity states to the THG spectrum can also be seen through the phase $\theta$ of the THG susceptibility: $\chi^{(3)}=|\chi^{(3)}|e^{i\theta}$.  In Fig.~5(b), the solid line represents $\theta$ obtained from the data of Eq.~(\ref{chi3}). Here $\theta$ has two hump structures at $\omega\sim2.3t$ and 3.2$t$.  The hump structure at $\omega\sim2.3t$ is caused by a three-photon resonance associated with the peak at the same energy in $\chi^{\rm THG}_{\parallel}$.  The position of the other hump structure at $\omega\sim3.2t$ is close to a two-photon resonance energy.  The phase obtained from the dominant term, Eq.~(\ref{THG}), is also plotted by the broken curve, which is similar to the solid curve.  The higher-energy hump structure disappears (as the dotted curve) the contribution from two-photon resonance is neglected in the denominator of Eq.~(\ref{THG}). 
\subsection{Cross-polarized TPA and THG}
The 18-site cluster employed here has two mirror planes associated with the $x$ and $y$ axes [see the inset of Fig.~1(a)] and thus have a $C_{ 2v}$ point group whose irreducible representation is $A_1$, $A_2$, $B_1$, and $B_2$.  Since the ground state of the Heisenberg model $\left|0\right>$ belongs to $A_1$, the photoexcited state with odd parity $\left|a\right>$ appearing in the dipole moment $x_{0a}$ ($y_{0a}$) belongs to $B_1$ ($B_2$).  On the other hand, the states with even parity $\left|b\right>$ encountered in $\chi^{\mathrm{THG}}_{\parallel}$ and $\chi^{\mathrm{TPA}}_{\parallel}$ belong to $A_1$.  In other words, the eigenstates of $B_1$ and $B_2$ representations contribute to both one- and three-photon resonance, and those of $A_1$ emerge as two-photon resonance.  Therefore, there is no information on $A_2$ subspace in $\chi^{\mathrm{TPA}}_{\parallel}$ and $\chi^{\mathrm{THG}}_{\parallel}$.  However, if we investigate cross-polarized nonlinear susceptibilities $\chi^{\mathrm{TPA}}_{\perp}$ and $\chi^{\mathrm{THG}}_{\perp}$, we are able to extract information about $A_2$ eigenstates because the state $\left|b\right>$ belongs to either $A_1$ or $A_2$.  In $\chi^{\mathrm{TPA}}_{\perp}$, the numerator of the dominant term, Eq.~(\ref{TPA}), becomes $x_{0a}y_{ab}(x_{bc}y_{c0}+y_{bc}x_{c0})+y_{0a}x_{ab}(y_{bc}x_{c0}+x_{bc}y_{c0})$, leading to two-photon resonance associated with $A_2$ subspace at $\Omega_b=2\omega$.  In contrast, the numerator of the dominant term of $\chi^{\mathrm{THG}}_{\bot}$ [Eq.~(\ref{THG})] is the sum of $ x_{0a}x_{ab} y_{bc}y_{c0}+y_{0a}y_{ab} x_{bc}x_{c0}$ and $x_{0a}y_{ab}(x_{bc}y_{c0}+y_{bc}x_{c0})+y_{0a}x_{ab}(y_{bc}x_{c0}+x_{bc}y_{c0})$, where two-photon resonance occurs within $A_1$ and $A_2$ subspaces, respectively.

In Fig.~3(a), $\chi^{\rm TPA}_{\bot}$ is plotted as the dotted curve.  As mentioned above, almost all of weight comes from the $A_2$ representation.   The weight is smaller than that of $\chi^{\mathrm{TPA}}_{\parallel}$ and a maximum appears in the higher-energy region around $\omega=4t$, although the weight spreads over in the energy range similar to $\chi^{\mathrm{TPA}}_{\parallel}$.  The $\chi^{\mathrm{THG}}_{\bot}$ spectrum is shown in Fig.~4 as the dashed curve.  Similar to $\chi^{\mathrm{THG}}_{\parallel}$, the contribution from three-photon resonance is dominated as evidenced by the energy position of the weight.  The $A_1$ contribution of two-photon resonance overlaps the three-photon resonance contribution and thus is hidden as a background.  The $A_2$ contribution, on the other hand, should appear around $\omega=4t$ according to the distribution in TPA, but the weight is not enhanced.  The total weight is smaller than that of $\chi^{\mathrm{THG}}_{\parallel}$, being similar to the TPA spectra.  In order to obtain further insight into the photoexcited states of the 2D Mott insulators, it is desired to measure experimentally the weight and distribution of the cross-polarized TPA and THG and compare them with our results.

\section{Summary}
We have examined the photoexcited states and nonlinear optical responses in the 2D Mott insulators by using an effective model for a single-band Hubbard model in the strong-coupling limit.  By employing the numerical technique for small clusters, we have obtained characteristic features that are consistent with the experimental data of linear absorption and TPA for 2D insulating cuprates.  They are the following:(i) linear absorption in 2D shows two broad peaks in contrast to a single peak in 1D, (ii) the magnitude of the TPA spectrum in 2D is smaller than that in 1D, and (iii) in 2D the peak and edge positions of the TPA spectrum are lower than those of linear absorption in contrast to  1D where TPA and linear absorption show almost an identical behavior near the edge.  The interplay of charge and spin degrees of freedom plays an essential role in such dimensionality dependence: In the photoexcited states of the 1D Mott insulator there is a separation of the two degrees of freedom, while in 2D the charge motion is strongly influenced by the presence of spins in the background.  Being consistent with this picture, we found that the features (i) and (iii) are sensitive to the value of the exchange interactions between spins.

In the THG spectrum of the 2D Mott insulators, dominant contributions are found to come from the process of three-photon resonance associated with the odd-parity states.  The two-photon resonance process is thus hidden by the dominant contributions.  The spectrum obtained by using realistic parameters for 2D insulating cuprates thus shows broad spectral features, which are qualitatively in agreement with the experimental data with a broad maximum,~\cite{Schumacher,Kishida3} but not to the data with pronounced peak structures.~\cite{Schulzgen}  We hope further the experimental efforts to clarify the characteristic of the THG spectrum in 2D.

We have calculated $\chi^{(3)}$ under the cross-polarized configuration of pump and probe beams.  It is shown in both the TPA and THG spectra that the spectral distribution is similar to $\chi^{(3)}$ with copolarized configuration but the weight is small.  Measurements of $\chi^{(3)}$ with the cross-polarized configurations are highly desired to obtain deep insight into the nature of photoexcited states in 2D Mott insulators.

\begin{acknowledgments}
The authors thank H. Okamoto and H. Kishida for valuable discussions.  This work was supported by a Grant-in-Aid for scientific Research from the Ministry of Education, Culture, Sports, Science and Technology of Japan, and CREST.  One of the authors (S.M.) acknowledges support of the Humboldt Foundation.  The numerical calculations were performed in the supercomputing facilities in ISSP, University of Tokyo, and IMR, Tohoku University.
\end{acknowledgments}


\begin{references}
\bibitem{Maekawa} See, for example, S. Maekawa and T. Tohyama, Rep. Prog. Phys. {\bf 64}, 383 (2001), and references therein.
\bibitem{Stephan} W. Stephan and K. Penc, Phys. Rev. B {\bf 54}, R17 269 (1996).
\bibitem{Gebhard} F. Gebhard, K. Bott, M. Scheidler, P. Thomas, and S. W. Koch, Philos. Mag. A {\bf 75}, 47 (1997).
\bibitem{Jeckelmann} E. Jeckelmann, F. Gebhard, and F. H. L. Essler, Phys. Rev. Lett. {\bf 85}, 3910 (2000).
\bibitem{Essler} F. H. L. Essler, F. Gebhard, and E. Jeckelmann, Phys. Rev. B {\bf 64}, 125119 (2001).

\bibitem{Mizuno} Y. Mizuno, K. Tsutsui, T. Tohyama, and S. Maekawa, Phys. Rev. B {\bf 62}, R4769 (2000).
\bibitem{Kishida1} H. Kishida, M. Ono, K. Miura, H. Okamoto, M. Izumi, T. Manako, M. Kawasaki, Y. Taguchi, Y. Tokura, T. Tohyama, K. Tsutsui, and S. Maekawa, Phys. Rev. Lett. {\bf 87}, 177401 (2001).
\bibitem{Butcher} See, for example, P. N. Butcher and D. Cotter, {\it The Elements of Nonlinear Optics} (Cambridge University Press, Cambridge, England 1990).
\bibitem{Kishida2} H. Kishida, H. Matsuzaki, H. Okamoto, T. Manabe, M. Yamashita, Y. Taguchi, and Y. Tokura, Nature, London, {\bf 405}, 929 (2000).
\bibitem{Ogasawara} T. Ogasawara, M. Ashida, N. Motoyama, H. Eisaki, S. Uchida, Y. Tokura, H. Ghosh, A. Shukla, S. Mazumdar, and M. Kuwata-Gonokami, Phys. Rev. Lett. {\bf 85}, 2204 (2000).
\bibitem{Tohyama} T. Tohyama and S. Maekawa, J. Lumin. {\bf 94-95}, 659 (2001).
\bibitem{Ashida} M. Ashida, Y. Taguchi, Y. Tokura, R. T. Clay, S. Mazumdar, Yu. P. Svirko, and M. Kuwata-Gonokami, Europhys. Lett., {\bf 58}, 455 (2002).
\bibitem{Schulzgen} A. Sch\"{u}lzgen, Y. Kawabe, E. Hanamura, A. Yamanaka, P.-A. Blanche, J. Lee, H. Sato, M. Naito, N. T. Dan, S. Uchida, Y. Tanabe, and N. Peyghambarian, Phys. Rev. Lett. {\bf 86}, 3164 (2001).
\bibitem{Schumacher} A. B. Schumacher, J. S. Dodge, M. A. Carnahan, R. A. Kaindl, D. S. Chemla, and L. L. Miller, Phys. Rev. Lett. {\bf 87}, 127006 (2001).
\bibitem{Kishida3} H. Kishida and H. Okamoto (unpublished).
\bibitem{Hanamura} E. Hanamura, N. T. Dan, and Y. Tanabe, J. Phys: Condens. Matter {\bf 12}, 8847 (2000).
\bibitem{Neudert} R. Neudert, M. Knupfer, M. S. Golden, J. Fink, W. Stephan, K. Penc, N. Motoyama, H. Eisaki, and S. Uchida, Phys. Rev. Lett. {\bf 81}, 657 (1998).
\bibitem{Soos} Z. G. Soos and S. Ramasesha, J. Chem. Phys. {\bf 90}, 1067 (1989).


\bibitem{Onodera} T. Tohyama, H. Onodera, K. Tsutsui, and S. Maekawa cond-mat/0205388.
\end{references}
\end{document}